
\baselineskip=10pt
\magnification=\magstephalf
\nopagenumbers
\parindent=3pc
\parskip=3pt
\font\sixteenrm=cmr17 at 16pt

\def\msun{M_{\odot}}

\def\etal{{\it et~al.\ }}
\def\eg{{\it e.g.,~}}
\def\ie{{\it i.e.,~}}
\def\de{\partial}
\def\ref{\noindent\hangindent.5in\hangafter=1}

\vskip 8truepc
\centerline{\sixteenrm  DYNAMICS OF CONDUCTIVE/COOLING FRONTS:}
\centerline{\sixteenrm  CLOUD IMPLOSION AND THERMAL SOLITONS.}
\vskip 5truepc
\centerline{\bf A. Ferrara$^{1,2,4}$ and Yu. Shchekinov$^{3}$}
\centerline{\it ${^1}$ Space Telescope Science Institute, 3700
San Martin Drive, Baltimore, 21218 MD, U.S.A.}
\centerline{\it ${^2}$ Osservatorio Astrofisico di Arcetri,
Largo E. Fermi, 5, 50125 Florence, Italy}
\centerline{\it ${^3}$ Institute of Physics, Rostov University,
194 Stachki, 344104 Rostov on Don, Russia}
\centerline{\it ${^4}$ Affiliated with the Space Science Department, ESA}
\vskip 5truecm
\centerline{\bf ABSTRACT}
\noindent We investigate the evolution of interfaces among
phases of the interstellar
medium with different temperature. It is found
that, for some initial conditions, the dynamical effects related to
conductive fronts are very important even if radiation losses, which
tend to decelerate the front propagation, are taken into account.
We have also explored the consequences of the inclusion of
shear and bulk viscosity,
and we have allowed for saturation of the kinetic effects. Numerical
simulations of a cloud immersed in a hot medium have been performed;
depending on the ratio of conductive to dynamical time, the density is
increased by a huge factor and the cloud may become optically thick.
Clouds that are highly compressed are able to stop the evaporation process
even if their initial size is smaller than the Field length.
In addition to the numerical approach, the time dependent evolution
has been studied also analytically.
Simple techniques have been applied to the problem in order to study
the transition stages to a stationary state.
The global properties
of the solution for static and steady fronts and useful relations among
the various physical variables are derived; a mechanical analogy is often
used to clarify the physics of the results. It is demonstrated that
a class of soliton-like solutions are admitted by the hydrodynamical
equations appropriate to describe the conduction/cooling fronts (in the
inviscid case) that do not require a heat flux at the boundaries.
Solitons are shown to result from the exact balance of convective and
conductive energy transport and we demonstrate that they can exist
only as associated with the fast velocity mode (analogous to the positive
Riemann invariant) of the system. Some astrophysical
consequences are indicated along with some possible applications to the
structure of the Galactic ISM and to extragalactic objects.
\vfill\eject
\centerline{\sl 1. INTRODUCTION}
\medskip
The problem of the thermal evaporation of cool gas clouds
embedded in a hot gas has received particular interest in
the last years. Apart from the richness of physical
aspects involved in the process, a strong astrophysical
motivation for the study of this type of systems is certainly provided by the
supposed ubiquitous presence of a hot phase of the
interstellar medium (ISM) in the galactic disk and halo, coexisting
with cooler phases.  It is therefore crucial to understand the
details of the physics to construct plausible models of the ISM.
\par
The first systematic approach to the problem of the effects of
thermal conduction on a cool cloud immersed in a hot medium
has been undertaken by Cowie \& McKee (1977). They obtained analytical
solutions  for the steady-state evaporation flow from the cloud; in
addition, they pointed out that the classical diffusion approximation
of  thermal conduction breaks down when the mean free path of electrons
becomes larger than the temperature scale (saturation). The relevance of
this effect
is quantified by the local parameter $\sigma_T$, which expresses the
ratio between classical and saturated conduction.
Some points are worth to be
mentioned: supposing that the flow is isobaric (or, equivalently,
that its  Mach number $M$ is small) implies that classical conduction
holds; secondly, if conduction is saturated, the usual hydrodynamic
equations are not valid anymore, due to  the large mean free paths
involved. Giuliani (1984) pointed out that their solutions are
obtained using a piece-wise method which leads to unphysical
discontinuities of the derivatives
of the variables at the transition points; his calculation
allows instead for a gradual change from the classical to saturated
regime.
\par
In a subsequent paper McKee \& Cowie (1977) generalized
the steady state problem to include
the effects of radiative losses, and they demonstrated the existence
of a critical length at which radiative losses balance the conductive
heating. A clear cut result for what concerns the steady state
of conductive/cooling (CC) flows is given by McKee \& Begelman (1990).
They recognized that the key parameter governing the structure of the
flow is the so-called Field length. This length corresponds to the
largest wavelength of a perturbation stabilized by thermal conduction
against thermal instability (Field 1965). Clouds smaller than the Field
length will suffer evaporation, while clouds larger than the Field length will
undergo evaporation if the pressure is lower than the saturated vapor
pressure (Penston
\& Brown 1970), otherwise they would undergo condensation. Remarkably
enough, this result is in agreement with the one found more than twenty
years before by Zel'dovich \& Pikelner (1969). These two studies, therefore,
clearly define the necessary conditions that  different ISM phases
must fulfill in order to coexist in a stable steady state.
\par
Aside from the consideration of radiative losses, a number of
physical processes may affect the evolution of conduction fronts.
Viscous stresses have been studied by Draine \& Giuliani (1984)
who show that they may have dramatic consequences on the
solutions, in particular when conduction is highly saturated. This
can be understood recalling that, if ion-electron temperature
equipartition holds, electron and ion mean free paths are almost
equal (Spitzer 1962). The main modifications to the results
obtained by Cowie \& McKee (1977) are a reduced Mach number of the
evaporative outflow and value of $\sigma_T$, and the appearance of
a collisionless shock transition. They also have allowed, in analogy with
saturated conduction, for viscosity saturation
when velocity gradients become large.
The impact of
conduction on complex multi-phase systems resembling the actual ISM has been
discussed by Balbus (1985) and Begelman and McKee (1990).
\par
As it can be realized from the above discussion, most of the
studies focused on the steady-state properties of the
evaporation/condensation process. A remarkable exception is
represented by the work of Doroshkevich and Zel'dovich (1981).
They have studied the temporal evolution of CC
fronts considering the effect of radiative losses, and have shown
that, if the hot gas is freely cooling, a cooling wave propagates
into the hot phase. Apart from the issue of the uncertain
existence of the intermediate asymptotic regime they explore
(for an extended discussion, see McKee \& Begelman 1990), their
thermal wave approach neglects all kind of hydrodynamic motions
which may arise in the system. One of the aims of this paper
is to demonstrate the relevance of such hydrodynamical effects
which, almost unavoidably, are related to the presence of
CC fronts, at least at the initial stages of
their evolution. The basic principle on which our convinction
is based is that, if the initial gradient of temperature at
the interface between the hot and cold phase is large enough,
the evaporation time $\tau_k$ becomes smaller than the cloud
characteristic dynamical time $\tau_d$. In this case the
pressure gradient, driving the motion, relaxes more slowly than the
thermal one. This importance of the dynamical aspects has been
emphasized also by Kovalenko \& Shchekinov (1992).
Thus, the interface behaves somewhat as an explosion
site, where the existing pressure excess generates an outward flow
(evaporation) and an inward one (implosion).
\par
In a very elegant paper Elphick, Regev \& Shaviv (1992) (but see
also Elphick, Regev \& Spiegel 1991) suggest a new approach
to the study of the dynamics of CC fronts. They
apply non-linear dynamics techniques, as the Lyapunov functionals,
to the investigation of the dynamics and interaction of a system
of localized structures constituted by the fronts in a thermally
bistable medium. In spite of
some approximations introduced into the problem (small thermal
conductivity, ideal cooling function), their approach reveals
extremely promising in the study of the dynamics of CC fronts.
In the second part of the present paper we also make use of
non-linear dynamics arguments to obtain the conditions under
which a particular class of hydrodynamical non-linear waves,
\ie soliton-like, may describe astrophysical CC fronts. Although
this subject can have an important impact on our understanding of
the structure of the ISM (Adams \& Fatuzzo 1992), it has not yet
been studied in the context of conductive flows. The present paper
will concentrate on the one dimensional case, but we feel that
higher dimensional studies may discover  an even larger  variety
of important physical aspects.
\par
The paper is organized as follows. In \S 2 we show the result of
numerical simulations where the effects of radiative losses,
viscosity, and ionization are included along, of course, with thermal
conduction. Saturation effects are also taken into account both for
conduction and viscosity.
In \S 3 we present some analytical
calculations that are intended to represent a slightly different
formulation of some of the aspects of the steady-state analysis, making use of
the dynamical ideas put forward by this paper; \S 4
deals with soliton-like solutions, and in \S 5 a brief summary
is given and some possible astrophysical consequences of the results
are discussed.


\bigskip
\centerline{\sl 2. EVOLUTION AND DYNAMICS OF CC FRONTS}
\medskip
Following the line presented in the Introduction, in this Section
we mainly want to demonstrate that under the conditions which are
thought to exist in the interstellar medium of galaxies, some
regimes may result in  strong non-linear, non-steady behaviour of the
system. Therefore, rather than explore the entire range of parameters
of the relevant physical quantities that may govern the
mass, momentum and energy exchange among the various phases of
interstellar medium, we concentrate on the cases that can
more appropriately show the kind of dynamical features that
are the main subject of the present work. Moreover, several
other  studies (see Introduction) have already investigated the static and
steady-state approximations and we refer the reader to that papers
for a general discussion.
\par
The actual interstellar medium is far from being homogeneous and
is characterized by the presence of strong gradients in the
hydrodynamical quantities at the interfaces among the various
phases by which it is constituted. Even when a pressure equilibrium
is achieved between spatially contiguous phases, those gradients
may induce bulk motions that eventually can drive the system away
from the initial equilibrium. The mass and energy exchange taking
place in these cases is governed, in general, by the combined
action of thermal conductivity, radiative energy losses, viscosity and
external heating provided to the system;
we can refer appropriately to it as a conductive/cooling (CC) front.
\par
In principle, a magnetic field may
also play a non-secondary role in the evolution of CC fronts.
However, as Balbus (1986) and Borkowski, Balbus
\& Fristrom (1990) have demonstrated, unless the magnetic field is
perpendicular to the front propagation direction, it is not likely to
appreciably modify the general evolution. On the other hand, if
magnetic field is  tangled and threads both hot and cold phases, the
reduction of the conductive transport coefficient is not well known;
some works indicate that this effect can be indeed a minor one, at
least for low densities (Rosner \& Tucker 1989; Tribble 1989).
For these reasons, we will neglect magnetic fields in the following;
the limits of this assumption will be discussed in \S~2.1.
\par

\par
The basic equations describing mass, momentum and energy
conservation can be set in the following form:
$${\de \rho \over \de t}=-\nabla \cdot (\rho {\bf v}),\eqno(2.1)$$
$$\rho {D{\bf v}\over Dt}=-\nabla\cdot \Pi,\eqno(2.2)$$
$${R\over \mu}\rho T {Ds\over Dt}=-\rho {\cal L}(p,T,x) +
\nabla\cdot(\kappa\nabla T)
+(\Pi- p \delta_{ij}){\de v_i\over \de x_j},\eqno(2.3)$$
$$p={R\over\mu} \rho T,\eqno(2.4)$$
where D/Dt is the Lagrangian derivative. The gas is supposed to be
perfect and with solar abundances, with temperature $T$, velocity $v$
and density $\rho$; the pressure $p$ is obtained from the equation
of state (2.4); $s$ is the specific entropy; other symbols have usual
meaning.
\par
The equilibrium cooling
function ${\cal L}$ has the usual form
$$\rho {\cal L}(\rho,T)= n^2 \Lambda(x,T)-n{\cal H}(\rho,T),\eqno(2.5)$$
where $n$ is the total particle number density, $x$ is the fractional
ionization of the medium and $n\Lambda$ and ${\cal H}$ are the cooling
and heating rates per particle. We have calculated ${\cal L}$
from the rates for microscopic processes given by Dalgarno
\& Mc Cray (1972) and Black (1981) at low temperatures, while for the
optically thin plasma case we have used the results obtained by Raymond
\etal (1976).
\par
The function $\kappa$ is the thermal conduction coefficient governed
by atomic diffusion at low temperatures (below $10^4$~K) and by
electronic diffusion at high temperatures. The adopted form of $\kappa$ is
$$ \kappa (x,T)= x \kappa_e T^{5/2} + (1-x) \kappa_a T^{1/2},\eqno(2.6)$$
where  $\kappa_e=6.7\times 10^{-6}$ and $\kappa_a=3.5\times
10^{3}$ in cgs units.
The stress tensor $\Pi$ is
$$\Pi_{ij}=-\eta\epsilon_{ij} - \zeta \delta_{ij} \nabla\cdot{\bf v}
+ p \delta_{ij}.\eqno(2.7)$$
The non-diagonal components of $\Pi$ are
$$\epsilon_{ij}=\left( {\de v_i\over \de x_j} + {\de v_j\over \de x_i}
- {2\over 3}\delta_{ij}{\de v_l\over \de x_l}\right),\eqno(2.8)$$
where $\eta=(2/3) {\sl Pr} \kappa/ R$ is the viscosity coefficient
corresponding to shear motions with Prandtl number {\sl Pr}, whereas
$\zeta$ is the bulk viscosity.
\par
The system (2.1)-(2.4) can be put in a non-dimensional
form if the physical variables are appropriately scaled. This allows
not only to write the equations in a simpler form, but it provides
a first insight on the physical scales of the problem. A natural
scaling is obtained if the cooling time $\tau_c= [R T
/(\gamma -1)\mu n \Lambda]_0$ is taken as time unit  and lengths
are expressed in terms of $\ell_F^2= (\kappa T/ n^2 \Lambda)_0$,
where the subscript $0$ indicates that the quantities in parenthesis
must   be evaluated at appropriate values of the variables. The
scale length $\ell_F$ is usually named ``Field length''; this
length corresponds to the maximum wavelength of linear
perturbations stabilized by thermal conduction. An additional
characteristic time $\tau_k= [p \ell^2 /(\gamma-1)\kappa T]_0$,
where $\ell$ is a fiducial size of the system, directly related to
thermal conductivity, can be introduced.
\par
An important point is that thermal conduction and
viscosity are intrinsically kinetic processes and therefore it is
crucial to understand the limits of any hydrodynamical approach
dealing with such phenomena. This limits are set by the condition
that the mean free path of the species (\ie atoms or electrons)
providing the energy and momentum transport must be
larger than the  scale length of the system,
typically represented by the Field length for the cooler gas.
As pointed out by several papers (Cowie \& McKee 1977;
Balbus \& McKee 1982; McKee \& Begelman 1990) this can be
a quite common situation in the interstellar and intergalactic
medium.  A correct solution of the problem in this case would require
the use of Boltzmann equation: unfortunately, this is possible only
for simple cases where particular simplifications can be introduced
(Draine \& Giuliani 1984). A good phenomenological approximation is
provided by the expression introduced by Giuliani (1984) for the
thermal flux ${\bf q}$
$${\bf q}=-\kappa \left( \nabla T\over 1 + \sigma_T\right),\eqno(2.9)$$
where $\sigma_T$, known as the saturation parameter, is
$$\sigma_T={\vert \kappa \nabla T\vert\over q_{sat}},\eqno(2.10)$$
and
$$q_{sat}\sim \rho v_{th}^3,\eqno(2.11)$$
where $v_{th}$ is the mean velocity for a Maxwellian distribution
of the particle velocities. A similar relation holds also for the
saturated viscosity (Draine \& Giuliani 1984). In strict analogy
with conduction they make the following position for the saturated
viscosity coefficient
$$\eta_T={\eta\over (1+\sigma_v)},\eqno(2.12)$$
where
$$\sigma_v=2 \eta \left( {\epsilon_m\over \varphi_v p_i}\right).$$
In the previous equation the parameter $\epsilon_m$ represents the
largest positive eigenvalue of the matrix $\epsilon_{ij}$, $p_i$ is
the usual ion isotropic pressure and  $\varphi_v$, describing
the maximum temperature anisotropy, can be reasonably
taken equal to 2. We adopted the same treatment also for the bulk
viscosity coefficient.
\par
Before going on and describe the numerical solutions, we believe
it is instructive to have an insight of the relevant scales of the
problem.
\par
The characteristic lengths of the problem defined above
are related by  a very simple expression:
$$\tau_k=\left(\ell\over \ell_F\right)^2 \tau_c.\eqno(2.13)$$
Equation (2.13) describes a line separating the $\tau_c$--$\tau_k$
plane in two regions (Fig. 1): the upper one ($\tau_k > \tau_c$) where
cooling processes dominate over conductive ones (\ie condensation)
and the lower one ($\tau_k < \tau_c$), dominated by
thermal conduction (\ie evaporation). A common interpretation
of the previous statement is probably based on the implicit assumption
that the dynamical behaviour of the system can be neglected and the
interface between the two phases can be treated as stationary.
This is equivalent to impose that both $\tau_c$ and $\tau_k$ are
much shorter than the dynamical response time of the system, which
can be approximated as $\tau_d \sim (\ell/c_s)_0$. This characteristic
time has in effect a slightly more delicate definition when two
gas phases with huge temperature differences are present.
In this case, the actual value of $\tau_d$ depends
on the details of the geometry of the system, but since it is likely that the
response of the colder phase to external pressure
perturbations is slower, its dynamical time at least to provides an
upper limit to this quantity. An usual astrophysical
situation consists of a cold cloud, of size $\bar \ell$ and
internal sound speed
$\bar c_s$, embedded in a hot medium, with sound speed $c_s$.
To simplify our considerations, we suppose that the cloud
is initially in thermal equilibrium, ${\cal L}=0$.
If we substitute $\ell= \bar c_s \tau_d$ in eq. (2.13),  the
locus of the points satisfying the condition $\tau_c=\tau_d$
is
$$\tau_k=\left({\bar c_s \over \ell_F}\right)^2 \tau_c^3.\eqno(2.14)$$
This curve instersects the one of eq. (2.13) in the point
$\tau_c=\ell_F/\bar c_s$ and, since $\tau_c=\tau_k$ on the
curve (2.13), at the intersection point is $\tau_k=\tau_c=\tau_d$
(Fig. 1). In order to identify the regions of the $\tau_c$--$\tau_k$ plane
corresponding to a different  $\tau_k/\tau_d$ ratio, we substitute
$\ell= \bar c_s \tau_d$ in eq. (2.13), with the additional condition
$\tau_d=\tau_k$. These relations provide
the locus of the points corresponding to $\tau_k=\tau_d$:
$$\tau_k=\left({\ell_F \over \bar c_s}\right)^2 \tau_c^{-1}.\eqno(2.14a)$$
These three curves divide the $\tau_c$--$\tau_k$ plane into six regions
with different ratios of the three characteristic times $\tau_c$, $\tau_k$,
and $\tau_d$, as shown in Fig. 1.
Therefore, once the temperature $T$ of the hot medium has been fixed,
four different regimes are possible (if
we neglect the patologic point at the intersection of the three curves),
actually depending on $\bar \ell$: this cases are labeled in Fig. 1
as A, B, C and D.
\par
Let us consider initially the subregions of this plane
for which the relation between the characteristic times are described by
strong inequalities ($\ll$ or $\gg$), represented by the dashed subregions
of the $\tau_c$--$\tau_k$ plane.
For a cloud with parameters corresponding to one of these subregions, the
initial dynamics could be described in terms of the possible evolutionary
modes (\ie isochoric, isobaric, isothermal).
However,
this description is not complete because, as discussed above,
three different times $\tau_c$, $\tau_d$ , $\tau_k$ govern the behaviour
of the system, and yet it is possible to have
different intermediate asymptotic regimes  sharing a similar initial behaviour.
Thus, in order to classify different
regimes of the evolution of a cloud embedded in a hot intercloud gas
it is necessary to take into account both the
initial behaviour and intermediate asymptotics.
We will assume, therefore,  that during the initial stages
of the evolution the relation between the largest characteristic times
is not changed by the different dependence
of those times on the dynamical variables. This holds for the sufficiently
deep parts of the dashed regions of $\tau_c$--$\tau_k$ plane. The main features
of the cloud evolution (initially in thermal equilibrium, as mentioned above)
can be summarized as follows:
\par\noindent
{\it Region 1} ($\tau_k \ll \tau_c \ll \tau_d$): initial isochorical
heating with subsequent cooling of the heated gas  at $\tau_c <t
\ll \tau_d$ --- no evaporation;
\par\noindent
{\it Region 2} ($\tau_k \ll \tau_d \ll \tau_c$): isochorical heating with
subsequent expansion of the heated gas at $\tau_d <t \ll \tau_c$ ---
evaporation;
\par\noindent
{\it Region 3} ($\tau_d \ll \tau_k \ll \tau_c$): isobarical expansion of the
slowly heated gas --- evaporation;
\par\noindent
{\it Region 4} ($\tau_d \ll \tau_c \ll \tau_k$): isobarical accretion of the
cooling intercloud gas close to the interface onto the cloud --- condensation;
\par\noindent
{\it Region 5} ($\tau_c \ll \tau_d \ll \tau_k$): isochorical cooling of the
intercloud gas with
subsequent slow accretion at $\tau_d <t \ll \tau_d$ --- condensation;
\par\noindent
{\it Region 6} ($\tau_c \ll \tau_k \ll \tau_d$): isochorical cooling of
intercloud
gas with subsequent slow accretion --- condensation.
\par
A straight consequence is that only the initial evolution of a cloud with
parameters lying in extreme regions of the $\tau_c$--$\tau_k$ plane (dashed in
Fig. 1) can be
considered as static or steady. The
hydrodynamical motions which develop in the intermediate stages
require a dynamical description.
This statement is even stronger if the cloud initial parameters are located
well into the undashed parts of Fig. 1, where it is impossible to
separate the
initial stages of the evolution from the intermediate asymptotics.
\par
The main aim of this paper is to investigate these dynamical effects
occurring at the interface among different phases of the ISM.
\vskip 1truecm
\centerline{\it 2.1 Numerical results}
\smallskip
In the following we present the results concerning the numerical
solution of eqs. (2.1)-(2.4).
A number of simplifying assumptions have been made to solve the
problem and we want to state them clearly to allow the reader to
evaluate the limitations of our results.
\par
We assume a) a spherical cloud immersed in a hot, homogeneous
surrounding medium; b) initial pressure equilibrium $nT=\bar n \bar T$,
where $n$ and $T$ are the density
and temperature of the hot gas, respectively, and barred quantities
refer to the cloud gas; c) the cloud is supposed to be initially in
thermal equilibrium (${\cal L}=0$) in an isobarically stable
state; the hot medium is instead allowed to cool down.  In practice,
given the ratio of the characteristic relevant times for the cases of
interest here, the temperature of the hot medium is almost constant
during the run; d) the ionization fraction is calculated at every time
step by equating the collisional ionization and recombination rates.
e) magnetic fields and non-equilibrium effects in the cooling function
are neglected.
\par
The main difficulty when solving the
system (2.1)-(2.4) consists in the large difference
among the time scales of the various processes. This is a typical
situation encountered in stiff problems, for which appropriate codes must be
developed. For our purposes we have adapted the numerical code
described by Kovalenko (1993). This is an implicit conservative
code in Lagrangian variables that conserves mass, energy and, in absence
of dissipation, entropy as well; the conservation of entropy
means that the effects of scheme viscosity are negligible.
\par
Fig. 2 shows the evolution of the relevant hydrodynamical variables
describing the cloud-hot gas system as a function of radius.
For sake of clarity, the displayed variables are normalized with
the corresponding cloud ones and the velocity is normalized with
the cloud internal sound speed $\bar c_s$.
The initial conditions are the following: cloud temperature $\bar T=
10^4$~K, cloud density $\bar n=1$~cm$^{-3}$, hot gas temperature
$T=3\times 10^6$; the cloud size is $\bar \ell=3.68$~pc.
{}From the pressure equilibrium condition, thus,
the hot gas density is $n=3.3\times 10^{-3}$~cm$^{-3}$.
This set of parameters  has been
chosen because they identify a C-type case according to the
notation introduced in Fig. 1. As already pointed out, these
cases are the most suitable one to elucidate the dynamical effects we are
interested in; by the way, this particular initial condition
is also a realistic one as far as the Galactic ISM is
concerned. Values of the pressure of this order ($p/k=10^4$~K~
cm$^{-3}$) are estimated in the Galactic disk, not only for the
thermal component but also for the magnetic field and cosmic ray
ones (Spitzer 1990). In addition, observational arguments
indicate that most of the HI mass in the thin disk is contained
in clouds of sizes 1-10~pc (Lockman, Hobbs \& Shull 1986).
The Field length for such a system is rather large: $\ell_F=1.249
\times 10^3~{\rm pc}\gg {\bar \ell}$, as expected for a dynamical evaporation
regime. It is also useful to compare the three relevant time scales
for this numerical experiment: $\tau_c=3\times 10^7$~yr, $\tau_k=8.6
\times 10^{-6} \tau_c$, $\tau_d=0.01 \tau_c$. Under these conditions
the electron mean free-path is $\lambda_e=11.9$~pc and therefore
thermal conduction is saturated; the global saturation parameter
(McKee \& Cowie 1977) is $\sigma_0 \sim 2 \lambda_e/{\bar \ell} =3.2$.
The saturation of thermal conduction is included in the
hydrodynamical code in the approximation described by eq. (2.9).
As discussed in the Introduction, there are reasons to
believe that when thermal conduction becomes saturated, this
is the case also for viscosity. For sake of comparison, however,
we present two numerical experiments that explore the cases
$\varphi=\infty$ (``classical'' viscosity) and $\varphi=2$
(saturated viscosity).
\par
Six different evolutionary stages, relative to the case
$\varphi=\infty$, are reported in Fig. 2.
At the
beginning the strong temperature gradient present at the
interface produces a huge localized thermal energy injection.
In some sense, the interface behaves as an explosion site and
the problem has some analogies, for example, with point
explosions. However, the explosion point is not fixed in space, but it
is comoving with the thermal front towards $r=0$. As a result of
the energy injection, two pressure perturbations are generated,
propagating in opposite directions, as illustrated by the
velocity profiles. The wave travelling with negative velocity
becomes supersonic already at about $t=0.15 \tau_d$, while the
one travelling away from the cloud surface remains subsonic
for all the subsequent evolution. This difference can be
easily understood recalling that $c_s /\bar c_s\sim
(T/\bar T)^{1/2} \sim 17$. The shoulder-like feature shown
by the temperature profile in the first three frames has
the same width of the overpressured region between the
two pressure discontinuities. This feature is due to the
density contrast between the left and the right boundary
regions: radiative losses are enhanced where the density is higher
and therefore the whole zone is strongly radiatively cooled.
Another interesting phenomenon is represented by the
compression of the cloud gas (for which $T \le 10^4$~K).
The spherical compression wave travelling towards the cloud center
increases the density of the cool material and, at the same time, it is
decelerated by the pressure gradient building up at the cloud
center. Therefore a core is formed with characteristic density
$\sim 100$ times the initial one.
We will discuss this point in more detail below.
When the pressure gradient inside the cloud has reached a critical
value, the implosion is stopped and the velocity changes
sign, \ie the wave is reflected. This is evident from the
fifth panel of Fig. 2, where also a pressure spike, which
will later relax driving the reflected wave, is
shown. The reflected wave will eventually superpose to the
initial outward wave, which at the time $t=\tau_d$ has
already reached $r \sim 3$. Note that the internal pressure
is one order of magnitude larger than at $t=0$.
We have also followed the evolution for $t>\tau_d$ which
can be summarized as follows. The mass loss in the outflow is
reduced to very small values; the
system settles down in a quasi-steady state,
if one neglects the damped small oscillations
of the interface around its equilibrium position, occurring
with a period of the order of
$\Delta t = 0.25 \tau_d$. These oscillations are the remnant
of the initial bouncing and they are the result of the
``elasticity'' of the system.
Eventually, at $t \sim \tau_c= 100 \tau_d$, the hot medium will
start to cool considerably and a cooling wave will propagate into
the ambient medium.
\par
An important result is that even if the cloud size is smaller than
the Field length, the cloud will not be completely evaporated in
cases like the present one (grouped under type C in Fig. 1) because
of the strong compression taking place in its internal regions.
In fact, the density and pressure increase induced by the compression
enhances the radiative losses, thus decreasing, and eventually reversing,
the front propagation velocity. A similar retarding effect due to cooling
losses can be realized also from an inspection of the isobaric case,
as we will show in \S~3.1, and it is responsible for the transition
to the steady state. However, the inclusion of the dynamics dramatically
amplifies its consequences. In other words, the non-linear effects
related to the dynamics (\ie compression waves) modify the ratio
between conductive heat flux and cooling, ultimately changing the Field length.
\par
It is also instructive to look at the evolution of the local
saturation parameter $\sigma_T$ at various radii. In Fig. 3
the behaviour of $\sigma_T$ is shown at three different evolutionary
times. Clearly, inside the cloud the conduction is always not saturated.
In the interface zone, instead, $\sigma_T$ can be as high as $\sim 100$,
but at $t=0.6 \tau_d$ it has already decreased to $\sim 8$ and
after that moment the conduction rapidly becomes classical. The peak
values of $\sigma_T$ are reached closely behind the interface,
where the temperature gradients are larger; however, secondary
structures appear (\ie the peak at larger radius for $t= 0.6
\tau_d$ and the dip for $t=\tau_d$) that are mainly governed
by the density changes corresponding to pressure jumps
(see eqs. [2.10] and [2.11]).
\par
Fig. 4 shows the evolution for the same parameters adopted
in Fig. 2, but now allowing for saturated viscosity
($\varphi=2$). The conclusions discussed in the previous
case still hold qualitatively. The main difference
is represented by the larger velocity of the outward wave
caused by the decreased efficiency of the viscous dissipation
of the pressure gradient. Therefore, being saturation of viscosity
much higher in the hot medium than in the cloud,
the gas compression
is much lower ($\sim 10$). Saturated viscosity, however,
does not seem to provide a mechanism capable to improve  the
cloud evaporation efficiency considerably.
\par
For completeness, we have performed several other runs for different
initial conditions. On this basis we can state that the numerical experiment
described before
has proven particularly exemplifying, as long as C-type initial
states are considered. In other words, maintaining approximately the same
value for the ratio $\tau_d/\tau_c$, but changing the temperature
of the hot medium, the results are qualitatively the same.
\par
The cloud implosion process we have shown deserves some
more detailed analysis given its astrophysical relevance.
To provide a more intuitive tool to the discussion, we
have reported in Fig. 5 (Plate 1) the evolutionary 2-D images of
the cloud for the same case presented in Fig. 3. The use
of density isocontours is aimed to point out the implosion
effect. From Fig. 5 the creation of a central dense core is
quite evident. The maximum value of central compression is
reached at $t=0.7 \tau_d$, when $\bar n(0)=253$~cm$^{-3}$
and the cloud size is 1~pc; the temperature in the cloud is in the range
$1.8\times 10^3 < \bar T < 2.5 \times 10^3$~K.
Initially the cloud has a mass of 5~$\msun$, and at
$t=0.7 \tau_d$ it has decreased to only 0.19~$\msun$. This corresponds
to an average  mass flux from the interface of about
$1.7 \times 10^{-6}$~$\msun$~yr$^{-1}$. After this moment,
however, the rate of mass loss is rapidly quenched, as already
discussed, and the cloud is able to survive, even if with a much
lower residual mass.
\par
Another consequence of the implosion process may
be of some relevance. The initial column density of the
gas for the case presented is $N_H=1.13\times 10^{19}$~cm$^{-2}$;
at the moment of maximum compression $N_H$ is increased
by a factor $7.2$. The cloud can become optically thick
and, if some dust is present, molecule formation may take place
in the interior shielded from energetic radiation.
\par
We turn now to the discussion of the possible model limitations
arising from the neglection of the magnetic field, $B$,
and non-equilibrium cooling. As far as $B$ is
concerned, we can estimate its dynamical influence in the following way.
When $B=0$, the front propagation is stopped, as seen above, at
the radius $r_m$ at which the internal pressure of the cloud has
reached a value $p_m$, roughly equal to the pressure
excess caused by conductive heating in the interface. The location
of this point is determined by the specific characteristics of the
dynamical evolution and energy balance of the front; in the example
studied  here, $r_m\simeq 0.2$. When a magnetic
field is present, a simple but meaningful way to extend the previous
analysis makes use of the virial theorem. For a spherical cloud, and
neglecting thermal pressure,  this is written as
$$4\pi r^3 p_m =  (1/3) r^3 B^2.$$
Using the flux-freezing condition $\pi r^2 B=\Phi=const.$, we obtain
the equilibrium radius $r_m$ for the magnetic (and static) case:
$$r_m=0.9\left(B_0^2\over 8\pi p_m\right)^{1/4}=0.9 \beta_0^{1/4},$$
where the subscript $0$ indicates initial values of the quantities.
Assuming equipartition between magnetic and thermal pressure, from
Fig. 2 we derive $\beta_0\sim 0.01$; therefore, in the magnetic case
$r_m=0.28$. According to this simple estimation, the main results
obtained here should not be appreciably modified by the inclusion
of $B$, because its effects would become important roughly
at the same stages as the thermal ones. This is not surprising if
the initial condition prescribes equipartition. Of course, these
arguments cannot substitute a much needed calculation in which
magnetic fields are included.  As already mentioned,
the only other possibility to suppress compression requires the
unlikely situation of a  perfectly insulated cloud such that thermal
conduction is almost completely inhibited.
\par
Non-equilibrium effects can certainly be important for the prediction
of radiative properties of the CC front. These effects have been studied
for example by Ballet \etal (1986) and Slavin (1989). However, looking
at Fig. 1 of Slavin (1989), in spite of large differences in the
cooling functions, the non-equilibrium temperature profiles obtained
differ very little from the equilibrium ones. This suggests that the
dynamics, mainly governed here by temperature gradients, should hardly
be affected. Thus, we believe that the equilibrium cooling function used
provides at least a reasonable first approximation for the study
tackled here.
\bigskip
\centerline{\sl 3. STEADY-STATE PROPERTIES}
\smallskip
The solutions discussed in the previous section illustrate the
richness of physical effects that may descend from the inclusion
of the dynamics in the study of CC fronts. Nevertheless,
the cases we have described may appear somehow ``extreme'',
in the sense that the cold phase is strongly affected and eroded by the
interaction with the hot gas.
Such violent dynamical effects are driven
by huge differences in the energy fluxes transported through the
system: these models, in fact, correspond to situations in which
heat input by thermal conduction and
radiative losses are strongly out of balance. This naturally leads
to the onset
of convective motions, as, for example, it is well known from the theory
of stellar interiors. In another formulation, according to the
mapping of the $\tau_c$--$\tau_k$ plane previously introduced, we
could say that for these states the various time scales are very
different and thus they are situated far from the bisectrix, where
$\tau_k=\tau_c$.
Although only a minority of the
real systems are so finely tuned to be located in the narrow region
of the plane closed to the bisectrix, they provide an interesting
sample of dynamical behaviours.
In addition, such a bisectrix may represent the final stage of the evolution
of a strongly dynamical system, after a transient period during which the
pressure gradient has relaxed.
\par
When the pressure can be considered almost constant, the stress tensor
reduces to $\Pi={\it const}$, and we are left with the energy
equation (2.3). This allows us to study the transition from the
last stages of the dynamical evolution into the steady
state. In the following we will derive a law for the
damped motion of the CC front based on simple flux balance principles.
The same approach  will prove to be quite powerful also
in order to understand the soliton-like solutions discussed below.
\vskip 1truecm
\centerline{\it 3.1 Slowering of the front due to cooling}
\smallskip
For $\Pi={\it const}$ motions are subsonic and the energy equation (2.3)
reduces to the standard heat equation modified by the presence of
cooling. To outline the main physical aspects, we study the simple planar
case, for which eq. (2.3) becomes
$${R\over (\gamma-1)\mu}\rho{\de T\over \de t}={\de \over \de x}\left[
\kappa {\de T\over \de x}\right] -\rho{\cal L}(p,T,x);$$
for isochorical processes the previous equation can be set in a
non-dimensional form  as described for the system (2.1)-(2.4)
$${\de T\over \de t}={\de \over \de x}\left[ \kappa {\de T\over \de x}\right]
-{\cal L}(p,T,x);\eqno(3.1)$$
to simplify the notation we have maintained for $\kappa/\rho$ the
same symbol $\kappa$.
In order to restrict the number of assumptions, we just take
${\cal H}=0$ and assume that $\Lambda$ does not depend on the
ionization, which is a good assumption for temperatures $> 10^4$~K.
To simplify the notation, we replace ${\cal L}(p,T,x)$ with $\lambda(T)$
in eq. (3.1). The problem then is completely defined when two
boundary conditions (BCs) are assigned along with the initial condition
(the problem is of parabolic nature)
$$\cases{T(-\infty,t)=0;\cr T(1,t)=1;\cr T(x,0)=0  &$x< 0$.\cr}$$
The discontinuity in the initial temperature profile will generate
a thermal front, whose position we define as the point $x_f$ at which
$T=0$,  propagating towards
$-\infty$. A numerical solution of eq.(3.1) with the cooling function
described in \S 2 is shown in Fig. 6. The effect of the cooling on
the dynamics of the front is evident, along with the deceleration of
the front to $\dot x_f=0$, which occurs at $t\sim 0.15$ and $x = -0.36$.
We interpret these results as the tendency
of the system towards the state in which conductive flux and volume energy
losses are balanced, thus leading to a truly stationary state.
In order to substantiate this statement we have performed
some additional analytical calculations.
\par
{}From a dimensional analysis of eq. (3.1) we obtain an
approximate expression for $x_f$ when
radiative losses are neglected:
$${T\over t} \simeq {T^{\beta+1}\over x_f^2},\eqno(3.2)$$
or $x_f\propto t^{1/2}$, having taken $\kappa\propto T^\beta$.
Note that the usual requirement for a fixed initial energy input (Zel'dovich
\& Raizer 1965) here has been released.
It is well known that this problem, with different BCs, admits
an exact self-similar solution  given by the
differential equation
$${d\over d\xi}\left(f^\beta {df\over d\xi}\right) + {\beta\over 2}
{df\over d\xi}=0.\eqno(3.3)$$
where $f$ and $\xi$ are appropriate self-similar quantities (Zel'dovich
\& Raizer 1965). As shown in \S 2, the inclusion of cooling
processes introduces different scaling laws.
An expression for the shape of the front can be obtained from
eq. (3.1), looking for solutions of the form $T=T(x-vt)$,
thus reducing the original partial differential equation to an
ODE. The solution, valid for ${\cal L}=0$, is
$$T(x)\sim\beta v \vert x_f -x\vert^{1/\beta}.\eqno(3.4)$$
To study the transition to a steady state we use flux balance arguments.
The energy input ${\cal E}_k$ due to thermal
conduction can now be dissipated by radiative energy losses ${\cal E}_c$.
Hence the front must decelerate and eventually be stop, for a certain
value $x_f^\star$ implicitly defined by
$$\vert {\cal E}_k\vert ={\de T\over \de t}= \lambda(T)= \vert
{\cal E}_c\vert;\eqno(3.5)$$
integrating eq. (3.5) over a column of gas of unit area
we have
$$\int_0^{x_f^\star}{\de T\over \de t}dx =
\int_0^{x_f^\star}\lambda(T)dx.\eqno(3.6)$$
The shape of non-linear conductive front (eq.[3.4]) is characterized
by an almost constant profile and a steep cutoff close to $x_f$. Therefore,
a good approximation, when evaluating the rhs integral, is to assume
$T={\it const}$, which gives
$$\vert {\cal E}_c \vert \sim x_f^\star.\eqno(3.7)$$
If we use eq. (3.4), and the relation $\dot x_f=(1/2)x_f^{-1}$, we can write
the lhs integrand of eq.(3.6) as
$${\de T\over \de t}={1\over 4x_f^2}(x_f-x)^{1\over \beta}\left[(x_f-
x)^{-1}-{\beta\over x_f}\right].\eqno(3.8)$$
It is easy then to show that
$$\vert {\cal E}_k\vert={1\over 4(1+\beta)} {x_f^\star}^{{1\over
\beta} -2},\eqno(3.9)$$
or, equating to eq. (3.7),
$$x_f^\star \sim \left[4(1+\beta)\right]^{\beta/(1-3\beta)}.
\eqno(3.10)$$
For electron conductivity, $\beta=5/2$ and $x_f^\star\sim 0.36$,
in excellent agreement with the numerical solution shown in Fig. 6.
\vskip 1truecm
\centerline  {\it 3.2 Static CC configurations}
\smallskip
As demonstrated above, the energy flux drive the system towards an equilibrium
state in which the energy input (due to
thermal conductivity) and output (due to radiative losses) are equal.
In a CC front, energy  is transported by conductive
$\bf q_\kappa=-\kappa \nabla T$ and convective
$\bf q_v=\rho\epsilon\bf v$ flux, where $\epsilon $ is the specific
thermal energy. In equilibrium must be
$$\oint (\bf q_v + \bf q_\kappa )d\bf S =\int \rho{\cal L}\cdot dV,
\eqno(3.11)$$
where the integration is performed in the rhs over the volume
of the system, and in the lhs over the surface delimitating the
volume, in the appropriate frame of reference in which ${\de/
\de t}=0$.
We shall consider an ensamble of scalar variables
$\wp$,  depending on $x$, $\wp=\wp(x)$, and velocity
${\bf v}=(v(x),0,0)$, $x\in (-\infty,+\infty)$.
\par
To illustrate our method we consider first the simplest case of  a static
medium, ${\bf v} =0$. This method is based on the local analysis of the
differential equations governing the system, and on the global properties
of the energy exchange which determine the characteristics of the flow at the
boundaries (\ie $\pm \infty$). A similar approach has been developed
recently by Elphick \etal (1991, 1992) who have investigated the interactions
of planar CC fronts. We will concentrate here on somewhat different aspects
of the problem, namely the general characteristics of the system for which
steady-state solutions of CC fronts do exist. For the particular
case ${\bf v}= 0$, the system (2.1)-(2.4) reduces to the equation
$${d\over dx}{\kappa(T){{dT} \over {dx}}}=\rho{\cal L}(\rho,T),\eqno(3.12)$$
with the BCs
$$\cases {T(-\infty)=T_1,\cr T(+\infty)= T_2;}\eqno(3.13)$$
the medium is assumed to be in thermal equilibrium at $x=\pm \infty$,
${\cal L}(\rho_1,T_1)={\cal L}(\rho_2,T_2)=0$.
\par
Since in the static case
$\rho (T) \propto 1/ T$ for isobaric and $\rho (T) ={\it const}$ for
isochoric case, respectively, we can re-write eq. (3.12) in a simpler form
$${d \over {dx}} {\kappa(T) {{dT} \over {dx}}}=\tilde{\cal L}(T).\eqno(3.14)$$
We define the non-dimensional variable $\theta={T /T_2}$ and the functions
$\kappa(\theta)= {\kappa(T) /\kappa(T_2)},$
$\tilde{\cal L}(\theta)={\rho{\cal L}(T) /[\rho_2 \kappa(T_2)
T_2]},$ and introduce the new variable $\eta(\theta)$ defined by
${d\eta}/{d\theta} = \kappa(\theta).$
Given the         univocal mapping $\eta \Leftrightarrow\theta$,
eq. (3.12) can be written also as
$${{d^2 \eta} \over{dx^2}} ={\tilde{\cal L}}(\eta).\eqno(3.15)$$
When $d{\tilde {\cal L}}/d \eta =0$, it must be $\vert d^2 {\tilde {\cal L}}
/d \eta^2\vert > 0$.
It follows that
$${dq^2\over{d\eta}} = 2{\tilde {\cal L}}(\eta),\eqno(3.16)$$
with $q\equiv {d\eta}/{dx}=0$ at the two points $\eta=\eta_1$ and $\eta=
\eta_2$. The existence of a static solution is therefore guaranteed
if the medium is thermally stable ($d{\tilde {\cal L}}
/d\eta > 0$) at $x=\pm\infty$ and $\int_{\eta_1}^{\eta_2} {\tilde {\cal L}}
d\eta=0$.
The last condition on the integral is equivalent to
eq. (3.11) and expresses the fact that cooling losses
are exactly balanced by heating.
There is an odd number of
stationary points (${\cal L} =0$) pertaining to thermally unstable
states with $d{\cal L}/{d\eta} < 0$; to simplify our analysis we consider
the case with  a single stationary point, $\eta_0$, for which, $d{\cal L}
/d \eta_0 < 0$.
\par
Near the i-th thermally stable point, the solution of eq. (3.15) can be
linearized
$${d^2\over{dx^2}} \delta\eta ={\tilde{\cal L}}_\eta (\eta)
\delta \eta,\eqno(3.17)$$
where $\delta \eta =\eta -\eta_i$, $\tilde{\cal L}_\eta={d\tilde{\cal L}}/
{d\eta} =(d\tilde{\cal L}/{d\theta})/\kappa$. Substituting
$\delta\eta \propto exp(\omega x)$ into eq. (3.17), we obtain
$$\omega =\pm \sqrt {\cal{~L}_\eta}.\eqno(3.18)$$
The previous relation describes a saddle point. The characteristic behaviour of
the phase trajectories near such a stationary point in the $q$-$\eta$
plane is plotted in Fig. 7a-b :
$$q \sim{ \pm {(\tilde{\cal L}_\eta)}^{1/2} \vert{\eta - \eta_i}\vert}, {\rm
{}~for~~} \tilde{\cal L}_\eta >0,$$
and
$$q \sim{(\vert\tilde{\cal L}_{\eta\eta}\vert/3)}^{1/2}\vert{\eta
-\eta_i}\vert^
{3/2}, {\rm~~for~~} \tilde{\cal L}_\eta=0.$$
A peculiar case corresponds to $\tilde{\cal L}(\eta) \sim \vert{\eta -
\eta_i}\vert^\alpha$ with $\alpha <1$, so that $\tilde{\cal L}_{\eta_i}=
+\infty.$
In this case the linearized equation (3.17) is not valid, and  a correct
solution can be obtained integrating directly eq.(3.16), which gives (Fig. 7c)
$$q \sim \pm \vert{\eta - \eta_i}\vert^{1+\alpha \over \alpha}.$$
\par
The thermally unstable point $\eta_0$ is a center, and
as a result, the phase diagram of the system described by eq.(3.12)-(3.13)
has the form plotted in Fig. 8a, and the eigenfunction of this system,
$T =T(x)$ (Fig. 8b), corresponds to the separatrix. It is easy to see that
the topology of the phase space allows for the existence of
a non-trivial solution ($T\ne const$) with $T(-\infty) =T(+\infty)=T_2$
(or $T_1$) , as shown in Fig. 9a-b. A final possibility
corresponds to a periodic structure
constituted by   clouds and intercloud gas  (Fig. 9c).
\par
To clarify the differences among these four cases we recur
to a mechanical analogy. We can re-write eq. (3.15)
in the following form:
$${{d^2\eta} \over{dx^2}}= -\nabla_\eta \Phi (\eta),\eqno(3.19)$$
where $\nabla_\eta \equiv \de / {\de \eta}$; $\Phi (\eta)=
-\int_{\eta_1}^{\eta}{\tilde{\cal L}}(\eta^\prime) d\eta^\prime$
can be seen as a ``potential''.
In the framework of this analogy, the characteristic size of the region with
$T\sim{T_2}$ is just the ``time interval'' $\Delta x$ needed for the mechanical
system (3.19) to make a complete orbit around the turning point $\eta =\eta_2$:
$$\Delta x (\eta_2) \sim \int_{\eta^{\prime}}^{\eta_2}{{d\eta}
\over {q(\eta)}},\eqno(3.20)$$
where $\eta^{\prime}>\eta_1$, for example $\eta^{\prime} =\eta_0.$
{}From eq. (3.16), the integral is equal to
$$\Delta x(\eta_2) \sim \int_{\eta^{\prime}}^{\eta_2}{{d\eta}\over
{\sqrt{2\vert{\Phi(\eta)}\vert}}};\eqno(3.21)$$
hence, $\Delta x(\eta_2)$ is determined by the behaviour of
$\Phi(\eta)$ near $\eta =\eta_2$. Three different possibilities can be
envisaged:
\par
\item{ a)} Let $\tilde{\cal L}(\eta )=\tilde{\cal L}_\eta(\eta -\eta_2) +
\tilde{\cal L}_{\eta\eta} (\eta -\eta_2)^2/2+...$, with finite values of
$\tilde{\cal L}_\eta, \tilde{\cal L}_ {\eta\eta},...$ In this case
$\Phi(\eta)\sim (\eta - \eta_2)^2 +...$ and the integral (3.21) has
a logarithmic singularity:
$$\Delta{x(\eta_2 -\epsilon) \sim \ln \epsilon},~~~{\rm for~} \epsilon
\rightarrow 0.$$
This corresponds to the solution shown in Fig. 8a-b.
\par
\item{ b)} $\tilde{\cal L}_\eta(\eta_2) =0, {\tilde{\cal L}_{\eta\eta}
(\eta_2)}\ne 0;$
$\Phi(\eta)\sim (\eta - \eta_2 )^3 +...$ and the integral (3.21)
has a power-like discontinuity at $\eta=\eta_2$:
$$\Delta{x(\eta_2 -\epsilon)}\sim \epsilon^{-1/2}.$$
Qualitatively, the profile of $T(x)$ is the same as in the previous case.
\par
\item{ c)} ${\cal~L}(\eta)\sim{-(\eta_2-\eta)^\alpha},\alpha <1,$ so that
${\cal~L}_\eta(\eta_2)=\infty;\Phi(\eta)\sim{U_0(\eta_2-\eta)^
{\alpha+1}}.$
The integral (3.21) has a finite value, and the $T(x)$ profile
is similar to the one plotted in Fig. 9.
\par
Thus, static CC fronts corresponding to a i)
finite cloud surrounded by hot intercloud gas, ii) finite
intercloud layer separating two infinite cold regions, are possible.
In addition, a configuration constituted by  iii) a sequence of flaky
cloud-intercloud
regions is also  possible if the cooling function $\tilde{\cal L}(\eta)$
has a singular behaviour near the stationary points $\eta_1$ and $\eta_2$.
These are the only solutions for which at least one of the two gas
components has a finite size.
Note that this requires ${\tilde{\cal L}_\eta}\ne 0$
and $\kappa(\eta_i)=0$ at the stationary points.
\par
We conclude that
the existence and the structure of steady-state CC fronts is determined
by the global relation (3.11) and by the local properties of the cooling
function
and of the conductivity: eq.(3.11)  expresses the
balance of the different mechanisms of energy transport through the system,
whereas $\tilde{\cal L}(T)$ and $\kappa(T)$ determine the behaviour of
the temperature near the stationary (saddle) points. An important feature is
that,  when $\kappa(T_i) =0$, the thermal conductivity is able to support the
energy transport near $T=T_2$ only over a finite interval $\Delta x$.
In Appendix A we clarify the relationship between this conclusion and the
Field length.

\vskip 1.0truecm
\centerline  {\it 3.3 Steady-state CC fronts}
\smallskip
The variety of solutions describing steady-state CC fronts with non-zero
velocity is, of course, richer than the static one. This is due to
the fact that, in general, the interaction of hot and cold gas is
likely to generate evaporation or condensation flows.
Convective motions, in addition, lead to a symmetry breaking
because, even if the temperature gradient is the same as in the static
case, the upstream heat flux differs from the downstream one.
In terms of the phase space $(q,\eta)$, this means that a convective heat flux
changes  the topology of the plane, and the velocity
can be seen as a bifurcation parameter.
Starting again from the integral equation (3.11) we want to obtain the
relation between the velocity and the global temperature and density
distribution for the steady-state case.
\par
Let us consider a steady-state flow corresponding
to an evaporation front travelling with velocity $u_0$. All the hydrodynamical
variables can be written in the form $a=a(x+u_0 t)$. In the frame
of reference comoving with the evaporation wave the appropriate equations are:
$$\nabla (\rho u) =0,\eqno(3.22)$$
$$\rho u \nabla u + \nabla p=0,\eqno(3.23)$$
$${R \over{\gamma -1}}{\rho u\nabla T}=-p\nabla u +\nabla(\kappa \nabla
T) -{\cal L}(\rho, T),\eqno(3.24)$$
$$p= {R\over \mu} \rho T.\eqno(3.25)$$
Here $u$ is the velocity in the comoving frame reference, $\nabla =d/dX$,
$X=x + u_0 t$. We introduce the following non-dimensional variables
$$\cases{\xi=X/\ell_{F_1}, ~~~~~\rho=\rho/ \rho_1, ~T=T/T_1, \cr
p=\gamma p/(\rho_1 c_1^2), U=u/c_1, ~\lambda=\ell_F {\cal L}/(\gamma
c_1^3 \rho_1),}\eqno(3.26)$$
where the subscript 1 refers to the unperturbed state at $X=-\infty$,
and $c_1=\sqrt {R \gamma T_1/\mu}.$
\par
The first integrals of motion are the following:
$$\rho U=U_0=J, \eqno(3.27)$$
$$\rho U^2 + p=U_0^2+1,\eqno(3.28)$$
$${\gamma \over{\gamma-1}} (pv-1)+{1 \over 2}
(U^2-U_0^2)=-{\lambda_s(\xi) \over U_0}, \eqno(3.29)$$
where $J=U_0$ is the mass flux, $v=\rho^{-1}$ is the
specific volume, and
$$\lambda_s(\xi)=\int_{-\infty}^{\xi} \lambda (\rho(\xi),T(\xi))
d\xi.\eqno(3.30)$$
Writing this system for $\xi= +\infty$, and adding the equation of state
$$\lambda (\rho, T)=0 \eqno(3.31)$$
for the gas at $\xi=+\infty$, we can obtain the values of
$p, \rho$, $U$ at $\xi=+\infty$; $U_0$
at $\xi=-\infty$ is equal to the velocity of the evaporation front for given
values of $\rho_1,p_1, T_1$ at $\xi=-\infty$ and $T_2$ at $\xi=+\infty.$
Note that equation (3.31) is not valid in the intermediate  range
$-\infty < \xi < +\infty.$
\par
Let us assume that $\lambda(\rho,T)=0$ has a shape similar to a
van der Waals-type equation,  shown in Fig. 10, usually
appropriate for a two-phase ISM model   (Field, Goldsmith \&
Habing, 1969). We stress, however, that {\it the results obtained in
this Section do not depend on the specific form of the cooling function
adopted but only on the number of its zeros}.
Thus,
$$p=\cases{(c_1^2 / \gamma) \rho~~{\rm for}~~ \xi=-\infty;\cr
(c_2^2 / \gamma) \rho~~ {\rm for}~~ \xi=+\infty,}$$
where $c_2 \gg c_1$ for the differences between
warm $(T_2 \simeq 10^4 K)$ and cold $(T_1 \simeq 10^2 K)$, or hot
$(T_2 \simeq 10^6 K)$ and warm $(T_1 \simeq 10^4 K)$ ISM phases.
In non-dimensional form the  equation of state for the hot gas
$(\xi=+\infty)$ is $$p_2=c_2^2 v_2^{-1}.\eqno(3.32)$$
The dynamical integrals of motion (3.27)-(3.29) can be reduced to
$$p_2=1+\gamma U_0^2 (1-v_2),\eqno(3.33)$$
and
$${\gamma \over {\gamma -1}} (p_2 v_2 -1)+{1 \over 2} (1-p_2) v_2-
{\gamma \over 2} U_0^2 (1-v_2) =Q,\eqno(3.34)$$
where $Q \equiv -\lambda_s/{U_0}, \lambda_s=\lambda_s(+\infty).$
Equation (3.33) is sometimes referred to as the Mikhelson equation (Zel'dovich
\& Kompaneets 1955),
while equation (3.34) is similar to a Hugoniot adiabatic for
$\lambda_s = 0$. Combining equations (3.33) and (3.34) with the equation
of state for the hot gas (3.32), we obtain
$${c_2^2 \over v_2} =1 + \gamma U_0^2 (1-v_2),\eqno(3.35)$$
and
$${\gamma \over{\gamma -1}} (c_2^2-1)+{1 \over 2} (1-{c_2^2 \over v_2})v_2
-{{\gamma U_0^2}\over 2} (1-v_2)=Q.\eqno(3.36)$$
Equation (3.35) has a solution for $\gamma U_0^2 \ll c_2^{-2}$
$$v_2=c_2^2(1+\gamma U_0^2 c_2^2 + o(\gamma U_0^2 c_2^2));\eqno(3.37)$$
after substitution into eq.(3.27) we obtain
$$U_2=c_2^2 U_0 (1+\gamma U_0^2 c_2^2+ o(\gamma U_0 c_2^2)).\eqno(3.38)$$
The assumed approximation $\gamma U_0^2 \ll c_2^{-2}$
corresponds to a subsonic flow at $\xi=+\infty$.
Under such conditions eq.(3.36) can be reduced to
$${{2 \gamma}\over{\gamma -1}} c_2^2 U_0 +\gamma U_0^3 c_2^4=
-2 \lambda_s,\eqno(3.39)$$
whose         solution is
$$U_0\simeq -{{\gamma-1}\over \gamma} {\lambda_s\over c_2^2}.
\eqno(3.40)$$
For $\lambda_s >0$ eq. (3.40) describes an
evaporation wave (net heating, $U_0 >0$),
whereas for $\lambda_s <0$ it describes a condensation wave
(net cooling, $U_0 <0$).
It can be shown (Appendix B) that this represents a subsonic flow at
$\xi=+\infty$ in accordance with the assumption made above.
\par
We now concentrate on the properties of the phase plane (Hayashi 1985 provides
an excellent background on this topic). It has to be
pointed out that the flow considered is approximately isobaric, as  derived
from eq. (3.40):
$$p = \rho c^2 = 1-\gamma U_0^2 c^2 +o(\gamma U_0^2 c^2)\eqno(3.41)$$
This leads to the following form of the energy equation (3.24)
$${1\over {\gamma-1}} U_0 {dT\over d\xi}+{dU\over d\xi} -
{d \over d\xi} \kappa {dT\over d\xi}=-\lambda(\rho,T).\eqno(3.42)$$
Combining the mass and momentum integrals, (3.27)-(3.28), with the equation of
state (3.25) we find
$$ T=-U^2 +{{1+U_0^2}\over U_0} U, \eqno(3.43)$$
which gives an estimation of the temperature valid for subsonic flows
$$ T\simeq {U \over U_0}.$$
After substitution into eq. (3.42), we can re-write the same equation as
$${d^2 \eta\over d\xi^2} -{\gamma\over{\gamma-1}} U_0{d\eta
\over d\xi}=\lambda (\rho,\eta),\eqno(3.44)$$
In analogy with the static case, we will assume that $\lambda(\rho,\eta)$
has three stationary points (Fig. 11a): two of them
correspond to thermally stable states ($\xi=-\infty$ and $\xi=+\infty$)
whereas the third ($\xi=\xi_0$) corresponds to a thermally unstable one.
For small
perturbations around the stationary points we have:
$${d^2 \delta\eta\over d\xi^2}-{\gamma \over {\gamma-1}} {U_0\over
\kappa} {d\delta\eta\over d\xi}=
\lambda_\rho \delta \rho + \lambda_\eta \delta \eta,\eqno(3.45)$$
where $\delta \rho \simeq -\rho^2 \delta \eta /\kappa$ for
isobaric perturbations. Taking
$\delta \eta \propto exp(\omega \xi)$ the characteristic equation
corresponding to (3.45) is
$$\omega^2 - {\gamma \over{\gamma-1}}  {U_0\over \kappa} \omega+
\left({\rho^2\over \kappa }\lambda_\rho-\lambda_\eta \right)
=0,\eqno(3.46)$$
and the eigenvalues are
$$\omega_{1,2} ={1\over 2} {\gamma\over {\gamma-1}} {U_0\over \kappa}
\pm \sqrt{{1\over 4}{\gamma^2\over {(\gamma-1)^2}}{U_0^2\over \kappa^2}-
{\left({\rho^2\over \kappa}\lambda_\rho -\lambda_\eta\right)}}.$$
Since for a thermally stable gas
${\rho^2/ \kappa}\lambda_\rho-\lambda_\eta <0,$
it is easy to see that the nature of the thermally stable points
corresponds to a saddle point (Fig. 11b). At the intermediate point $\eta_0$,
instead, ${\rho^2/ \kappa} \lambda_\rho - \lambda_\eta >0,$
When condition  (3.40)  holds, the discriminant is negative because $U_0^2
=o((\ell_F/\ell_c)^2 c_2^{-4})$, whereas $U_0=o((\ell_F/\ell_c) c_2^{-2})$ and
$\lambda_\rho,\lambda_\eta = o(\ell_F/\ell_c).$ Thus $\eta_0$ is an unstable
spiral point (Fig. 11b), and the phase diagram differs qualitatively from
the static case. The main difference is that evaporation  (Fig. 11c)
and condensation (Fig. 11d) waves are not
symmetric anymore.  Thus, a steady-state CC front can exist
only as a switch-on wave between two stable phases
(cold/hot), and periodic structures like those plotted in Fig. 9c
are not realisable.
\par
To better understand this difference we apply again the mechanical
analogy used above. Equation (3.44) can be written as follows:
$${d^2\eta\over d\xi^2} -{\gamma\over {\gamma-1}}{U_0\over \kappa}
{d\eta\over d\xi}=-\nabla_\eta \Phi(\eta).\eqno(3.47)$$
In this case the potential $\Phi(\eta)$ is an asymmetric function of $\eta$
(Fig. 12a-b); therefore, for the evaporative case ($d\eta/ d\xi >0$) a negative
``friction force'', like the one represented by the second term on the lhs of
eq. (3.50), increases the ``energy'' of the ``particle''
during its motion
from $\eta=\eta_1$ to $\eta=\eta_2$. Since $\Phi(\eta_1)=0$,  we have
$\Phi(\eta_2) >\Phi(\eta_1)$ for an evaporation wave,
and $\Phi(\eta_2) < \Phi(\eta_1)$ for a condensation wave. This means that
it is impossible for this ``particle'' to reach $\eta=\eta_2$ from
$\eta=\eta_1$ and turn back, because such transitions lead to a difference
between the initial and final values of the ``potential'' at $\eta=\eta_1.$
Out of analogy, this is due, of course, to the radiative losses occurred
in the gas during the phase transitions.
The eigensolution of eq. (3.47) corresponds to the equality between
the ``potential'' barrier and the ``work'' made by the negative
``friction force''.
Indeed, multiplying eq. (3.47) by $d\eta/ d\xi$ and integrating
over $\xi(-\infty,+\infty)$ with the BCs $(d\eta/ d\xi)_{\pm\infty} =0$,
we obtain
$$\int_{\eta_1}^{\eta_2}{\gamma\over {\gamma-1}}{{U_0}\over \kappa}\left({
{d\eta}\over {d\xi}}\right)^2 d\eta =\cases {\Phi(\eta_2)-\Phi(\eta_1),
\hbox{~~~for   evaporation};\cr \Phi(\eta_1)-\Phi(\eta_2), \hbox{~~~for
condensation}.\cr}$$
\par
An artificial way to maintain a steady-state CC front with a structure of a hot
(cold) gaseous layer surrounded by cold (hot) gas (similar to that of the
static CC configuration plotted in Fig. 9)    is to inject a fixed heat flux
at a certain $\xi$. In such a way, the
additional heat flux behind the condensation wave provides heating
to the cooled gas and evaporates it, restoring the initial state (Fig. 13a).
The specific value of the heat flux necessary to maintain such a CC front is
determined by the value of $d\eta/ d\xi$ at the top of the phase
trajectory coming out of the critical point $\eta=\eta_2$ (in Fig. 13a it
corresponds to the point S). In the case of an evaporation wave, a negative
heat flux behind the CC front dilutes the excess of thermal energy, provides
a cooling mechanism to the hot gas and favours the condensation process
(Fig. 13b). However, this requires that the heat source  moves together
with the CC front.
\bigskip
\centerline{\sl 4. EXISTENCE OF SOLITON-LIKE SOLUTIONS}
\smallskip
The previous results demonstrate that a CC front structure similar to a
soliton (condensation - evaporation [CE] wave or evaporation - condensation
[EC] wave) is possible if supported by an artificial heat source.
However, the hydrodynamical integrals contain non-linear terms that may
cause the
same effect. According to eq. (3.43)
$$\Delta T=-2 U \Delta U +{{1+U_0^2}\over U_0} \Delta U.$$
The two terms on the rhs produce a perturbation of $T$ in opposite
directions. This perturbation is small only if $\vert U_0\vert \ll 1$,
as in the static case. Thus, we
expect that the non-linear term $U^2$ in eq. (3.43)
may result in a qualitative change of the phase plane topology that
allows soliton-like solutions without any artificial heat source at the
boundary.
\par
Let us consider the energy equation (3.24), describing a
stationary wave with velocity  $U_0$
$${U_0\over {\gamma-1}}{dT\over d\xi}=-p {dU\over d\xi} +
{d\over d\xi}\kappa{dT\over d\xi}-\lambda(\rho,T),\eqno(4.1)$$
and with BCs $T(-\infty)=T(+\infty)=T_1.$ Note that $U_0$ in eq.
(4.1) is in general different from the $U_0$ determined in the previous Section
for evaporation or condensation waves. Using the relations among $T$, $p$
and $U$ determined by the first integrals we have
$$p = U_0^2 + 1 -U_0  U = \epsilon_0 - U_0  U,\eqno(4.2)$$
$$T = -U^2 + {{1+U_0^2}\over U_0} U = -U^2 +{\epsilon_0\over U_0} U,
\eqno(4.3)$$
where $\epsilon_0 = 1+U_0^2$ is the specific enthalpy. Equation (4.3)
allows us to reduce eq. (4.1) to a form containing only $T$.
For this purpose we solve eq. (4.3) to find $U = U(T)$:
$$U ={1\over 2}{\epsilon_0\over U_0}\left(1 \pm \sqrt{1-{{4 U_0^2}\over
\epsilon_0^2} T}\right) \equiv f^{\pm} (T),\eqno(4.4)$$
where $\pm$ correspond to the ``fast'' and ``slow'' velocity mode,
respectively.  After
substitution of eq. (4.2) into eq. (4.1), using equation (4.4), we can
re-write eq. (4.1) as follows
$${d^2T\over d\zeta^2} -{U_0\over 2} \left({{\gamma+1}\over {\gamma-1}}
- F^\pm \right){dT\over d\zeta}-\kappa \lambda (\rho (T),T) =0,\eqno(4.5)$$
where $d\zeta = d\xi /\kappa$, and
$$F^\pm =\pm \left(1-{{4 U_0^2}\over \epsilon_0^2}T \right)^{-1/2}.$$
\par
We further assume that the cooling function has two equilibrium points:
$T_1 =1$, and $T_2$ defined by $\lambda(\rho,T) = 0$. These two points are
stationary points on the phase plane $T$-$(dT/ d\zeta)$, that are $(T_1,0)$
and $(T_2,0)$. The solution corresponding to a solitary (EC or CE) wave is
possible if one of the points is a center and the other one is a saddle point.
The characteristic equation
for linear perturbations $\delta T \propto \exp (\omega \zeta)$
for the two stationary points is
$$\omega^2 -{U_0\over 2} \left({{\gamma +1}\over {\gamma-1}} - F_i^\pm\right)
\omega -\left({dL\over dT}\right)_i =0,\eqno(4.6)$$
where $i$ =1,2 and $dL/ dT = \kappa(T_i)(d\lambda
/dT)_i$. Hence, the eigenvalues are
$$\omega_{1,2}^{(i)} ={U_0\over 4} \left({{\gamma+1}\over {\gamma-1}} -
F_i^\pm\right)
\pm \sqrt{{U_0\over 16}^2 \left({{\gamma+1}\over {\gamma-1}} - F_i^\pm\right)
^2 + \left({dL\over dT}\right)_i}.\eqno(4.7)$$
The thermally stable point with $(dL/dT)_i >0$
is a saddle point, whereas the nature of the thermally unstable point
with $(dL/dT)_i <0$ depends on the value
of $\left({{\gamma+1}/{\gamma-1}} - F_i^\pm \right).$ We analyze now in detail
the case that corresponding to soliton-like solutions which are
the subject of this Section.
To  be specific, we suppose that the saddle point
is identified with $T = T_1$, and $T = T_2 > T_1$ is the center. The phase
diagram of the system in this case is plotted in Fig. 14a; the relative
temperature eigenfunction is shown in Fig. 14b.
\par
It is easy to see that the point $(T_2,0)$ can be a center only for
the fast mode solution:
$$U = {1\over 2}{\epsilon_0\over U_0} \left(1 + \sqrt{1 -{{4 U_0^2}\over
\epsilon_0^2} T}\right)\equiv f^+,\eqno(4.8)$$
with the necessary condition ${{\gamma+1}\over {\gamma-1}} - F^+ =0$,
equivalent to the equation:
$${1\over {\sqrt{1 - {{4 U_0^2}\over \epsilon_0^2} T_2}}} = {{\gamma+1}\over
{\gamma-1}},\eqno(4.9)$$
whose solution is
$$U_0^2 = -\left[1-{{(\gamma+1)^2}\over 2 \gamma} c_2^2\right]
\pm \sqrt{\left(1-{{(\gamma+1)^2}\over 2 \gamma}
c_2^2\right)^2-1}. \eqno(4.10)$$
This solution determines two values of $U_0^2 >0$ if the sound
speed at the  center point satisfies the inequality $c_2^2 >4\gamma/
(\gamma+1)^2$, valid for $T_2 >T_1=1$.  In this case,
there exist two different soliton-like solutions corresponding
to supersonic and subsonic solitary waves of EC type.
The FWHM of the soliton can be estimated directly from eq.(4.5).
At the top of the soliton $dT/d\zeta=0$ and $d^2T/d\zeta^2<0$, $\kappa
\lambda \ne 0$, hence
$$\Delta x\sim \sqrt{{{\kappa \lambda} \over{\vert d^2T/d\zeta^2\vert}}}.$$
All quantities should be taken for $T=T_\star$, where $T_\star$ is the
temperature at the soliton top (or bottom for a CE soliton);
$T_\star$ can be estimated at the point where $\Phi =0$, ($\Phi$ is
defined by eq. [3.19]).
\par
A similar analysis for the case with a saddle point at $T=T_2$ (thermally
stable phase) and a center point at $T=T_1$ demonstrates the
existence of a CE soliton-like solution with a reversed shape with
respect to the EC case.
\par
These soliton-like solutions correspond to the separatrix of
the phase plane (see Fig. 14a) like the hydrodynamical solitons
described by the Korteweg-de Vries equation. However, their physical nature
is quite different. In fact, they are the result of the simultaneous
action of four processes:
thermal conductivity $\nabla(\kappa \nabla T)$, the cooling ${\cal L}
(\rho,T)$, convective transport $\rho {\bf u} \nabla T$ and advection
$p\nabla {\bf u}$. Indeed, the
thermal conductivity determines the second-order characteristic equation (4.6)
and the set of solutions near the stationary points; the radiative cooling
determines the number of stationary points and the local topology of the phase
space in their vicinity. The most interesting role is played by the last two
processes: their conspiracy is able to restore the symmetry of the phase
space broken by the hydrodynamical motions for slow isobaric flows
(\eg eq. [3.33]). The sum of the hydrodynamical terms in eq. (4.1),
taking into account eqs. (4.2) and (4.3), can be written as:
$${U_0\over {\gamma-1}}{dT\over d\xi} + p{dU\over d\xi} =
{U_0\over 2}\left({{\gamma+1}\over {\gamma-1}} + {\epsilon_0\over U_0}
{dU\over dT}\right) {dT\over d\xi}.$$
Thus, the first term on the rhs represents an asymmetric
heat flux and hence destroys the symmetry of the phase space present in
the isobaric case ($dU/dT =0$). The second term describes a stabilization or
amplification of the convective heat flux depending on the sign of
$dU/dT$. For the fast mode
$dU/dT < 0$, thus the convective heat flux is stabilized and balanced for
an appropriate value of $U_0$ at a certain $T=T_1$ (or $T=T_2$).
This provides the soliton-like shape of the wave. The soliton temperature
profile, however, is slightly asymmetric as it can be realized by
rewriting eq. (4.5) in the following form
$$q{dq\over dT}-{U_0\over 2}\left({{\gamma+1}\over{\gamma-1}}-F^\pm\right)
q-\kappa \lambda=0.$$
It is clear that the upper and lower parts of the ($T,q$) plane are
asymmetric: at the points where ${dq \over d\zeta }=0$, in fact,
$T(q)$ for $q>0$ differs from $T(q)$ for $q<0$.

\bigskip
\centerline{\sl 5. SUMMARY AND POSSIBLE APPLICATIONS}
\smallskip
In this paper we have studied the dynamics of thermal conduction
fronts in a multi-phase medium with parameters suitable to describe
the general ISM. The inclusion of radiative losses affects
substantially both the dynamics and the structure of the conductive/cooling
front. However, a number of additional physical effects
like viscosity, ionization, saturation of the kinetic processes have
been shown to be able to modify appreciably the overall behaviour
of the system, when included. The presence of several, often
very different, scales in the problem determines regimes in
which the response of the system is strongly time dependent and
non-linear.
CC fronts may induce relevant dynamical effects if
the conductive time is much shorter that the sound crossing time of the
cloud. This condition is, in practice, equivalent to require that
the cloud size is smaller than the Field length for the hot gas.
\par
As an example, we have explored in detail a realistic dynamical regime by
means of numerical simulations.
The adopted initial condition consists of a spherical cloud
in equilibrium with a surrounding hot medium at a pressure
$p/k=10^4$~cm$^{-3}$~$K$, with a ratio of intercloud-to-cloud gas temperature
of $T/\bar T =300$. This set-up may fairly well reproduce a typical
interstellar Galactic environment. The main results are summarized as
follows:
\item{ 1.} For the case $\varphi=\infty$ (classical viscosity) a huge
pressure gradient is created at the interface hot/cool gas which
drives an outward and an inward flow. In some sense, the interface
behaves like an explosion site. The supersonic wave travelling towards
the cloud center compresses the cloud and at the same time tends to
evaporate it. A dense ($> 100$ times the initial density) core is formed;
the subsequent increase in pressure is able to stop the compression
wave and to reflect it back.
\par
\item{ 2.} We have shown that after the pressure gradient has relaxed,
the front is decelerated, and  the system
tends to a steady-state. Using simplifying positions as neglecting the
viscosity and thermal conduction saturation, the final stages of this
process have been investigated exploiting flux balance arguments.
These findings have been substantiated by their agreement
with exact numerical solutions.
\item{ 3.} Clouds of size smaller than the Field length can be able eventually
to stop the evaporation process due to the density and pressure increase
following the compression phase; the net results is that the enhanced
radiative losses decelerate the CC front (non-linear effect).
\item{ 4.} In the first evolutionary stages, thermal conduction is highly
saturated with a value of $\sigma_T\sim 100$, but it rapidly becomes
classical after about one half of the dynamical time.
The role of the dynamics in decreasing the saturation level of the
kinetic effects is an important piece of information to add to the
results for the steady cases obtained by McKee \& Cowie (1977) and
Draine \& Giuliani (1984)
\item{ 5.} If saturated viscosity is allowed, $\varphi=2$, the compression
factor is decreased by a factor of $\sim 10$, but the main features
of the flow remain qualitatively the same.
\par
The global properties of static and steady-state CC front solutions have been
derived from an analysis of their behaviour in the phase plane.
Different thermal phases individuate stationary states whose
topological nature is determined by the stability properties of
those states. Depending on the behaviour of the cooling function at
the stationary points, it has been shown that in the static
case three different configurations are possible:
i) finite cloud surrounded by hot intercloud gas, ii) finite
intercloud layer separating two infinite cold regions, and
iii) a periodic sequence of cloud-intercloud regions. The last possibility
is related to the fact that in this case thermal conduction can
transport energy only over a finite spatial interval.
The addition of a  velocity field produces a symmetry
breaking in the topology of the solutions and, therefore, periodic
structures are not allowed.
\par
It has been shown that solitary wave solutions are admitted by
the set of equations describing CC fronts, even in the case of
no heat flux at the boundaries, thanks to the
intrinsic non-linear character of these equations.
Physically, those soliton-like solutions are the result of the
exact balance between
convective and conductive heat flux, which restores
the symmetry breaking occurring, in general, when the flow is not
isobaric. These solitary waves have the remarkable property that,
differently from an evaporation front that leaves the gas at
a higher temperature behind it, they change the temperature of the
medium affected by their passage but after this transient the gas
returns to its initial state.
\par
Finally, we add here that it is possible to demonstrate the existence
of self-similar solutions of the equations describing
CC fronts in absence of bulk motions. Their properties are currently
under investigation and they will be discussed elsewhere. The preliminary
results show that, contrary to the suggestion of
Doroshkevich \& Zel'dovich, cooling media do not necessarily induce
condensation waves, but they may excite evaporation modes in the
system.
\par
The most severe limitation of our model descends from the
spherical geometry  used in the numerical calculations. The main reason
for this choice is, of course, dictated by simplicity requirements;
real ISM clouds can largely deviate from this idealisation.
Some of the findings of this paper can be modified by  more
refined and multi-dimensional calculations; the striking nature of
the implosion effect, in particular, can be partially an artifact of the
geometry. However, we believe that the basic physics (compression,
density increase, slowering of the front, wave reflection) underlying the
process, and outlined throughout the paper, must be realized also by
more complex configurations.
In addition, this simplification highly facilitates the comparison
with analogous classical works in the field, based on the same assumption.
\par
Some mathematical problems related to the existence of solitons
need some more accurate analysis. For example, it is not clear to
us if the the non-linear terms present in the hydrodynamic equations
are likely to destroy closed trajectories in the phase space around the
center point for large amplitudes. In addition, the possibility of closed
paths seems to depend on the nature of the system, which in turn is
determined by
its characteristic equation (Guckenheimer \& Holmes 1983). Finally, the
stability of solitons for the Korteweg-de Vries equation has been
demonstrated  by Benjamin (1972), but the analogous question for the
thermal solitons we are presenting here is completely open.
\par
Although this paper is not particularly dedicated to a specific
astrophysical situation, a number of consequences worth to be
explored can be suggested.
\par
A result that has important and immediate implications
for our understanding of the ISM descends from the cloud implosion induced by
thermal conduction. Due to the unavoidable density increase,
the cloud tends to become
optically thick, providing a site for molecule formation. A clear evidence
of molecular clouds embedded in neutral hydrogen, both
surrounded by hot, x-ray emitting plasma, has been provided by the ROSAT
shadowing experiments (Burrows \& Mendenhall 1991; Snowden \etal 1991)
in the direction of the Draco complex.
The phenomenon
we have studied requires cloud properties which are by far typical of
the HI clouds populating the Galaxy. Therefore, it may take place
whenever very hot gas ($T\sim 3\times 10^6-10^7$~K) is suddenly
injected by some process (the most natural being supernova explosions)
in their surroundings.  In a couple of papers, Bertoldi (1989)
and Bertoldi \& McKee (1990) have discussed an analogous cloud compression
mechanism based on the action of an ionization-shock induced by
the radiation field of newly born stars. Our mechanism is rather
important in the regions in which the radiation field in not
particularly enhanced, but some hot gas is present. A
prototype of these regions can be the Galactic halo, where
hot gas is suggested by several different observations (Sembach
\& Savage 1992; for a review see Spitzer 1990) along with neutral HI clouds.
Those clouds can be a product of a thermal instability occurring
in a fountain flow under particular conditions (Houck \&
Bregman 1990, Ferrara \& Einaudi 1992, Li \& Ikeuchi 1992).
However, it has been pointed out (Ferrara \& Einaudi 1992) that
for the prevailing conditions of the hot flow, turbulence is
more easily generated than a condensation; in this case the
clouds should have a different origin.
If the predictions of the McKee \& Ostriker (1977) model are
correct, suitable conditions for the implosion mechanism can
easily be found in the disk as well.
\par
Some other aspects of the cloud/hot gas interaction deserve
additional study. For example, clouds engulfed by
a strongly anisotropic hot gas flow can be rocket-accelerated as
a whole, by the same mechanism providing
implosion in the spherically symmetric case. This may have
important consequences for the observed degree of HI
turbulence required by some models to support the extended
neutral Galactic component (Lockman \& Gehman 1991; Ferrara 1993).
Also, the interaction of a cloud with
the hot surrounding medium may induce turbulent motions
inside the cloud. These motions are likely to be supersonic, as
can be realized from the results of \S 2.1. Therefore, even in
presence of a substantial dissipation, the energy
flux due to conductivity may be able to sustain a remarkably high
degree of supersonic turbulence for a long time interval in the cloud.
\par
Finally, we would like to mention that soliton solutions may
lead to a very different picture of the interstellar medium.
The temperature fluctuations related to their wavy nature
can create time-dependent patterns which can propagate
and survive for long times due to their intrinsic stability.
This may have some relevance for the interfaces producing
the highly ionized species (C~IV, N~V and O~VI) detected either
in the disk and in the halo by absorption line measures.
The study of extragalactic environments where two-phase media
are found, (\eg protogalactic clouds in a hot intergalactic medium,
emitting clouds in the Broad Line Region of AGNs, cooling flows) constitutes
an appealing application of the theory developed here.
However, still more physical insight can be achieved by a dedicated analysis
in which the three dimensional properties of soliton solutions are
considered.
\vskip 3truecm
\centerline{\sl ACKNOWLEDGMENTS}
\smallskip
We are grateful to I. Kovalenko for his valuable help with the
numerical code; we thank S. Balbus, G. Field, E. Kasak,
C. Norman, P. Pietrini, E. Saar, P. Traat
for discussions, G. De Marchi for help with graphics, and the
Institute of Astrophysics of Tartu where the work has been completed.
YuS acknowledges the hospitality of the Arcetri Observatory.
This work has been partially supported (AF) by an ESA External Fellowship.

\vskip 2truecm
\centerline{\sl APPENDIX A}
\smallskip
The solutions found for static CC fronts are intimately connected
with the Field length.  To see this, we recall that eq. (3.14)
for $\kappa(T)=0$ has the form
$$\kappa_T\left({dT\over{dx}}\right)^2_{T=T_2}=\tilde{\cal L}(T_2)=0,$$
and $$\kappa_T\left({{d^2T}\over{dx^2}}\right)_{T=T_2}
=({\tilde{\cal L}_T})_{T_2}.$$
For $\kappa_T(T_2)\ne 0$ and $\tilde{\cal L}_T(T_2) \ne 0$, this
results in a finite size of the region with $T \simeq T_2$, of the order of
$$\Delta x(T_2)\sim \sqrt{{ T_2\kappa_T(T_2)}\over {\tilde{\cal L}_T(T_2)}}.$$
The fact that $\Delta x(T_2)$ comes out to be of the order of the Field
length $\ell_F$ is not surprising. This is in agreement with the
discussed interpretation of $\ell_F$ as the length over which
cooling and conductive energy fluxes are able to establish equilibrium.
\par
The previous statement can be analogously demonstrated also in the case
$\kappa(T) \ne 0$, in which conductivity provides heat transport over
the whole system.
At $\eta=\eta_0$ (center), eq.(3.16) has the following solution
$$q^2=q_0^2 -\vert{{\cal~L}_{\eta_0}}\vert (\eta -\eta_0)^2,$$
where $q_0^2 =2\int_{\eta_1}^{\eta_0}{\cal~L}(\eta) d\eta.$
It  follows that
$$\eta=\eta_0\pm q_0^2  \ell_{F_0}^2 \sin {x \over
{\lambda_{F_0}}},$$
with $\ell_{F_0} =\ell_F(\eta_0)$. Similarly, for
$\eta=\eta_2$ (saddle), the same equation has the solution
$$\eta=\eta_2 [1-e^{-x/\ell_{F_2}}],$$
with $\ell_{F_2} =\ell_F(\eta_2).$

\vskip 2truecm
\centerline{\sl APPENDIX B}
\smallskip
We show here that the flow described by eq. (3.39  is subsonic at
$\xi=+\infty$. First we recall that
$$\vert U_2\vert = \vert U_0 \vert v_2 \simeq {{\gamma-1}\over \gamma}
\vert \lambda_s \vert; $$
If $\vert\lambda_s\vert \sim \lambda$, then $\vert U_2\vert \ll c_2$ and the
flow is subsonic.
This can be shown the following way.
We can express the non-dimensional cooling function $\lambda$ in the
equivalent form
$$\lambda ={{\ell_F{\cal L}_m {\cal L}} \over{\gamma c_1^3 \rho_1
{\cal L}_m}} \sim {\ell_F \over \ell_c} \tilde \lambda,$$
where ${\cal L}_m \ne 0$ is the absolute value of the cooling function
at a certain temperature $T_1 < T < T_2$; for example, we can assume
that ${\cal L}_m$
is the maximum of $\vert {\cal L}\vert$ in the interval $(T_1,T_2)$; $\ell_c
=c_1 \tau_c$ is the wavelength whose inverse frequence is equal to the cooling
time; $\tilde \lambda ={\cal L}/{\cal L}_m$ is the cooling function
normalized to its maximal value ${\cal L}_m.$ The integral $\int_{-\infty}
^{+\infty} \tilde \lambda (T) d\zeta \sim \lambda$; this means that $\vert
\lambda_s \vert \sim \ell_F/\ell_c$. Taking into account that $\kappa \propto
v_T/(\sigma_{el} n)$, where $v_T \sim c_1$ is the thermal velocity,
$\sigma_{el}$ is the
cross section for elastic scattering of particles providing heat transfer, and
${\cal L} \propto (\sigma_{el} n v_T) \Delta E p_{in}$, where $\Delta E$ is
themean energy radiated by two colliding particles in inelastic processes,
and $p_{in}$ is the
mean probability for inelastic processes, we obtain $\ell_F/
\ell_c \sim (p_{in} \Delta E/k T_1)^{1/2}$. This estimate demonstrates that
$\ell_F/\ell_c$ is much smaller than unity because inelastic
processes are sufficiently less frequent than elastic ones. For example,
in a free-free radiating gas at $T \sim 10^6$~K,
$p_{in} \Delta E/k T \sim v_T/c$
where $c $ is the light speed (Landau \& Lifshits 1975).
These arguments demonstrate that evaporation or condensation in a stationary
flow are always subsonic.

\bigskip
\parindent=0pc
\parskip=6pt
\centerline{\sl REFERENCES}
\vskip 2pc
\ref Adams, F. C. \& Fatuzzo, M. 1992, Univ. of Michigan, preprint

\ref Balbus, S. A. 1985, ApJ,  291, 518

\ref Balbus, S. A. 1986, ApJ, 381, 137

\ref Balbus, S. A. \& McKee, C. F. 1982, ApJ, 252, 529

\ref Ballet, J., Arnaud, M. \& Rothenflug, R. 1986, A\&A, 161, 12

\ref Burrows, D. N. \& Mendenhall, J. A. 1991, Nature, 351, 629

\ref Begelman, M. C. \& McKee, C. F. 1990, ApJ, 358, 375

\ref Benjamin, A. 1972, Proc. Roy. Soc., A328, 153

\ref Bertoldi, F.  1989, ApJ, 346, 735

\ref Bertoldi, F. \& McKee, C. F. 1990, ApJ, 354, 529

\ref Black, J. H. 1981, MNRAS, 197, 555

\ref Borkowski, K.J., Balbus, S. A., \& Fristrom, C.C. 1990, ApJ, 355, 501

\ref Cowie L.L \& McKee, C. F. 1977a, ApJ, 211, 135

\ref Dalgarno, A. \& McCray, R. A. 1972, ARA\&A, 10, 375

\ref Doroshkevich, A. G. \& Zel'dovich, Ya. B. 1981, Sov. Phys. JETP, 53, 405

\ref Draine, B. T. \& Giuliani, J. L. 1984, ApJ, 281, 690

\ref Elphick, C., Regev, O. \& Spiegel, E. A. 1991, MNRAS, 250, 617

\ref Elphick, C., Regev, O. \& Shaviv N. 1992, ApJ, 392, 106

\ref Ferrara, A. \& Einaudi, G. 1992, ApJ, 395, 475

\ref Ferrara, A. 1993, ApJ, 407, 157

\ref Field, G. B. 1965, ApJ, 142, 531

\ref Field, G. B., Goldsmith D. W. \& Habing, H. J. 1969, ApJ, 155, L149

\ref Giuliani, J. L. 1984, ApJ, 277, 605

\ref Guckenheimer, J. \& Holmes, P. 1983, Nonlinear Oscillations, Dynamical
     Systems and Bifurcations of Vector Fields, (Berlin:Springer-Verlag)

\ref Hayashi, C. 1985, Nonlinear Oscillations in Physical Systems, (Princeton:
     Univ. Press)

\ref Houck, J. C. \& Bregman, J. N. 1990, ApJ, 352, 506

\ref Kovalenko, I. 1993, J. Comp. Phys., submitted

\ref Kovalenko, I. \& Shchekinov, Yu. 1992,  Astron. Astophyis. Trans., 1, 129

\ref Landau, L.D. \& Lifshits, E.M. 1975, The Classical Theory of Fields,
     (Oxford:Pergamon)

\ref Li, F. \& Ikeuchi, S.  1992,  ApJ, 390, 405

\ref Lockman, F. J., Hobbs, L. M. \& Shull, J. M. 1986, ApJ, 301, 380

\ref Lockman, F. J., \& Gehman, C. S. 1991, ApJ, 382, 182

\ref McKee, C. F. \& Cowie, L.L. 1977, ApJ, 215, 213

\ref McKee, C. F. \& Ostriker, J. P. 1977, ApJ, 218, 148

\ref McKee, C. F. \& Begelman, M. C. 1990, ApJ, 358, 392

\ref Penston, M. V. \& Brown, F. E. 1970, MNRAS, 150, 373

\ref Rosner, R. \& Tucker, W. H. 1989, ApJ 338, 761

\ref Raymond, J. C., Cox, D. P., \& Smith, B. W. 1976, ApJ, 204, 290

\ref Sedov, L. I. 1957, Similarity and Dimensional Methods in Mechanics,
     (New York: Academic)

\ref Slavin, J. D., 1989, ApJ, 346, 718

\ref Sembach, K. R. \& Savage, B. D.  1992, ApJSS, 83, 147

\ref Snowden, S. L., Mebold, U., Hirth, W., Herbstmeier, \& Schmitt,
     J. H. M. M., 1991, Science, 252, 1529

\ref Spitzer, L. 1962, Physics of Fully Ionized Gases, (New York:Interscience)

\ref Spitzer, L. 1990, ARA\&A, 28, 71

\ref Tribble, P. C. 1989, MNRAS, 238, 1247

\ref Zel'dovich, Ya. B. \& Kompaneets, A. S. 1955,  Theory of Detonation,
     Moscow

\ref Zel'dovich, Ya. B. \& Raizer 1965, Physics of Shock Waves and High
     Temperature Hydrodynamic Phenomena, (London:Academic Press)

\ref Zel'dovich, Ya. B. \& Pikelner, S.B. 1969, Sov. Phys. JETP, 29, 170
\vfill\eject

\bigskip
\centerline{\sl FIGURE CAPTIONS}
\smallskip
\vskip0.2in
{\bf Figure 1} Different possible regimes for an inhomogeneous medium
as a function of the ratio between the characteristic conductive and cooling
times. The linear curve corresponds to eq. (2.13), the cubic curve corresponds
to eq. (2.14), and the negative slope curve corresponds to eq.(2.14a);
the vertical line corresponds to a fixed temperature of
the hot phase; dashed regions correspond to states with large
differences in the values of the characteristic times.
\vskip0.1in
{\bf Figure 2} Evolutionary stages of a CC front as a function
of radius in units of $\bar \ell$ for the case $\varphi=\infty$.
The parameters adopted are $\bar n=1.0~{\rm cm}^{-3}$, $\bar
T=10^4~{\rm K}$, $\bar \ell=3.68~{\rm pc}$, $T=3\times 10^6~{\rm K}$.
In each panel the flow Mach number ({\it solid line}), the
density ({\it dotted}), and the {\it log} of temperature ({\it dashed})
and pressure ({\it long-dashed}),
all normalized to the cloud values, are reported. Note that in the last four
panels the density is divided by a factor of 10, as indicated
by the label $n/10$.
\vskip0.1in
{\bf Figure 3} Evolution of the local saturation parameter $\sigma_T$
as a function of the cloud radius for the same case shown in Fig. 2.
\vskip0.1in
{\bf Figure 4} The same as Fig. 2, but $\varphi=2$.
\vskip0.1in
{\bf Figure 5} The same as Fig. 2, density isocontours.
\vskip0.1in
{\bf Figure 6} Evolution of an isobaric thermal conduction front
(numbers correspond to time elapsed from $t=0$) with a real cooling
function. {\it Solid curves} full solution, {\it dashed curves} no cooling
included.
\vskip0.1in
{\bf Figure 7} Segments of phase trajectories near the stationary point
$\eta_1$ for a) $\tilde{\cal L}_{\eta_1} >0$, b) $\tilde{\cal L}_{\eta_1} =0$,
and c) $\tilde{\cal L}_{\eta_1} = +\infty$.
\vskip0.1in
{\bf Figure 8} a) Phase diagram corresponding to the system (3.12), (3.13)
; b) eigensolution of (3.12), (3.13).
\vskip0.1in
{\bf Figure 9} Phase trajectories and eigenfunctions of eq. (3.12) with
a) $T(-\infty) = T(+\infty) = T_2$  (cold cloud immersed in a hot
intercloud gas); b) $T(-\infty) = T(+\infty) = T_1$  hot
bubble in a cold infinite gas; c) periodic structure of cloud - intercloud gas.
\vskip0.1in
{\bf Figure 10} Van der Waals-type equation of state for interstellar gas.
The dotted lines correspond to the intermediate region $-\infty <\xi <+\infty$.
\vskip0.1in
{\bf Figure 11} a) Schematic dependence of the cooling function on $\xi$:
for an evaporation wave cooling function is asymmetric and
heating dominates ($\lambda_s <0$); b) structure of the integral curves near
the critical points; c) phase diagram for an evaporation CC front and d)
for a condensation  one.
\vskip0.1in
{\bf Figure 12} Asymmetric ``potential'' curve.  The straight
line shows the increase in potential energy due to a negative ``friction''.
\vskip0.1in
{\bf Figure 13} Special case of a solitary a) condensation-evaporation  and b)
evaporation-condensation wave with fixed conductive heat flux at the moving
boundary $S$.
\vskip0.1in
{\bf Figure 14} a) Phase diagram and b) soliton-like profile of T for the
CC front described by eq. (4.5) with ${\gamma+1}/{\gamma-1} = F(T_2)$;
the thermally stable phase is located at $T = T(\pm \infty) = T_1$.
\vfill\eject
\bye